\documentclass[runningheads]{llncs}
\usepackage{graphicx}
\usepackage{flushend}
\usepackage{latexsym}
\usepackage{makecell}
\usepackage{multirow}
\usepackage{enumitem}
\usepackage{todonotes}
\usepackage{subcaption,booktabs}
\usepackage{mathdots}
\usepackage{amsfonts}
\usepackage{amssymb}
\usepackage{bm}
\usepackage{calrsfs}
\usepackage[group-separator={,},group-minimum-digits={3}]{siunitx}
\usepackage{microtype}
\usepackage{multicol}
\usepackage[multiple]{footmisc}
\usepackage[hyperfootnotes=false]{hyperref}
\begin{document}
\title{\textsc{TabLeX}: A Benchmark Dataset for Structure and Content Information Extraction from Scientific Tables}
\titlerunning{TabLeX}

\author{ 
Harsh Desai\inst{1} \and
Pratik Kayal\inst{2} \and
Mayank Singh\inst{3}
}
\authorrunning{Desai \textit{et al.}}
%
\institute{Indian Institute of Technology, Gandhinagar, India \\ \email{hsd31196@gmail.com}
\and
Indian Institute of Technology, Gandhinagar, India \\ \email{pratik.kayal@iitgn.ac.in}\\
\and
Indian Institute of Technology, Gandhinagar, India  \\ \email{singh.mayank@iitgn.ac.in}}

\maketitle  
\begin{abstract}
Information Extraction (IE) from the tables present in scientific articles is challenging due to complicated tabular representations and complex embedded text. This paper presents \textsc{TabLeX}, a large-scale benchmark dataset comprising table images generated from scientific articles. \textsc{TabLeX} consists of two subsets, one for table structure extraction and the other for table content extraction. Each table image is accompanied by its corresponding \LaTeX~source code. To facilitate the development of robust table IE tools, \textsc{TabLeX} contains images in different aspect ratios and in a variety of fonts. Our analysis sheds light on the shortcomings of current state-of-the-art table extraction models and shows that they fail on even simple table images. Towards the end, we experiment with a transformer-based existing baseline to report performance scores. In contrast to the static benchmarks, we plan to augment this dataset with more complex and diverse tables at regular intervals. 

\end{abstract}
\section{Introduction}
\label{section1}

Tables are compact and convenient means of representing relational information present in diverse documents such as scientific papers, newspapers, invoices, product descriptions, and financial statements~\cite{tkasar}. Tables embedded in the scientific articles provide a natural way to present data in a structured manner~\cite{douglas1995using}. They occur in numerous variations, especially visually, such as with or without horizontal and vertical lines, spanning multiple columns or rows, non-standard spacing, alignment, and text formatting~\cite{embley}. Besides, we also witness diverse semantic structures and presentation formats, dense embedded text, and formatting complexity of the typesetting tools~\cite{singh2019automated}. These complex representations and formatting options lead to numerous challenges in automatic tabular information extraction (hereafter, \textit{`TIE'}). 

\noindent \textbf{Table detection vs. extraction:}
In contrast to table detection task (hereafter, \textit{`TD'}) that refers to identifying tabular region (e.g., finding a bounding box that encloses the table) in an document, TIE refers to two post identification tasks: (i) table structure recognition (hereafter, \textit{`TSR'}) and (ii) table content recognition (hereafter, \textit{`TCR'}). TSR refers to the extraction of structural information like rows and columns from the table, and TCR refers to content extraction that is embedded inside the tables. Figure 1 shows an example table image with its corresponding structure and content information in the \TeX~language. Note that, in this paper, we only focus on the two TIE tasks.

\noindent \textbf{Limitations in existing state-of-the-art TIE systems:} There are several state-of-the-art tools (Camelot\footnote{\url{https://github.com/camelot-dev/camelot}},Tabula\footnote{\url{https://github.com/chezou/tabula-py}}, PDFPlumber\footnote{\url{https://github.com/jsvine/pdfplumber}}, and Adobe Acrobat SDK\footnote{\url{https://www.adobe.com/devnet/acrobat/overview.html}}) for text-based TIE. On the contrary, Tesseract-based OCR~\cite{tesseract} is commercially available tool which can be used for image-based TIE. However, these tools perform poorly on the tables embedded in the scientific papers due to the complexity of tables in terms of spanning cells and presence of mathematical content.
The recent advancements in deep learning architectures (Graph Neural Networks~\cite{gnn} and Transformers~\cite{vaswani2017attention}) have played a pivotal role in developing the majority of the image-based TIE tools. 
We attribute the limitations in the current image-based TIE primarily due to the training datasets' insufficiency. Some of the critical issues with the training datasets can be (i) the size, (ii) diversity in the fonts, (iii) image resolution (in dpi), (iv) aspect ratios, and (v) image quality parameters (blur and contrast). 

\noindent \textbf{Our proposed benchmark dataset:} In this paper, we introduce \textsc{TabLeX},  a benchmark dataset for information extraction from tables embedded inside scientific documents compiled using \LaTeX-based typesetting tools. \textsc{TabLeX} is composed of two subsets---a table structure subset and a table content subset---to extract the structure and content information. The table structure subset contains more than three million images, whereas the table content subset contains over one million images.  Each tabular image accompanies its corresponding ground-truth program in \TeX~macro language. In contrast to the existing datasets~\cite{table2latex,pubtabnet,publaynet}, \textsc{TabLeX} comprises images with 12 different fonts and multiple aspect ratios. 

\noindent \textbf{Main contributions:} The main contributions of the paper are:
\begin{enumerate}[nosep,noitemsep]
    \item Robust preprocessing pipeline to process the scientific documents (created in \TeX~language) and extract the tabular spans. 
    \item A large-scale benchmark dataset, \textsc{TabLeX}, comprising more than three million images for structure recognition task and over one million images for the content recognition task.
    \item Inclusion of twelve font types and multiple aspect ratios during the dataset generation process. 
    \item Evaluation of state-of-the-art computer vision based baseline~\cite{feng2020scene} on \textsc{TabLeX}.
\end{enumerate}

\noindent \textbf{The paper outline :} We organize the paper into several sections. Section~\ref{sec:rel_work} reviews existing datasets and corresponding extraction methodologies. Section \ref{sec:dataset} describes the preprocessing pipeline and presents  \textsc{TabLeX} dataset statistics. Section~\ref{sec:metrics} details  three evaluation metrics. Section \ref{sec:exp} presents the the baseline and discusses the experimental results and insights. Finally, we conclude and identify the future scope in Section \ref{section6}.

\begin{figure}[h]
\centering
    \subfloat[\centering Table Image]{{\includegraphics[width=0.5\textwidth]{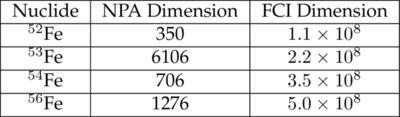} }}%
    \qquad
    \subfloat[\centering Structure Information ]{{\includegraphics[width=0.5\textwidth]{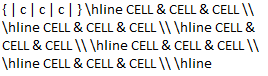} }}%
    \qquad
    \subfloat[\centering Content Information ]{{\includegraphics[width=0.5\textwidth]{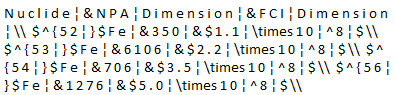} }}%
    \caption{Example of table image along with its structure and content information from the dataset. Tokens in content information are character-based, and the `¦' token acts as a delimiter to identify words out of a continuous stream of characters.}
    \label{fig:table_example}%
\end{figure}

\section{The Current State Of The Research}
\label{sec:rel_work}
In recent years, we witness a surge in the digital documents available online due to the high availability of communication facilities and large-scale internet infrastructure. In particular, the availability of scientific research papers that contain complex tabular structures has grown exponentially. However, we witness fewer research efforts to extract embedded tabular information automatically. Table~\ref{datasets} lists some of the popular datasets for TD and TIE tasks. Among the two tasks, there are very few datasets available for the TIE, specifically from scientific domains containing complex mathematical formulas and symbols. As the current paper primarily focuses on the TIE from scientific tables, we discuss some of the popular datasets and their limitations in this domain. 

\begin{table}[!t]
\caption{Datasets and methods used for Table Detection (TD), Table Structure
Recognition (TSR) and Table Recognition (TR). * represents scientific paper datasets.}
\label{datasets}
\centering
\resizebox{\hsize}{!}{
\begin{tabular}{|c|c|c|c|c|c|c|}
		\hline
        \multicolumn{1}{|c|}{\textbf{Datasets}} &
		\multicolumn{1}{|c}{\textbf{TD}} &
		\multicolumn{1}{|c}{\textbf{TSR}} &
        \multicolumn{1}{|c}{\textbf{TR}} &
        \multicolumn{1}{|c}{\textbf{Format}} &
        \multicolumn{1}{|c|}{\textbf{\# Tables}} &
        \multicolumn{1}{|c|}{\textbf{Methods}}\\ \hline
        Marmot\cite{marmot}* & \checkmark & $\times$ & $\times$ & PDF & 958 & Pdf2Table\cite{pdf2table},TableSeer\cite{tableseer},\cite{hao2016a}\\ \hline
        PubLayNet\cite{publaynet}* & \checkmark & $\times$ & $\times$ & PDF & 113k & F-RCNN\cite{frcnn}, M-RCNN\cite{maskrcnn}\\ \hline
        DeepFigures\cite{deepfigures}* & \checkmark & $\times$ & $\times$ & PDF & 1.4m & Deepfigures\cite{deepfigures}\\ \hline
        ICDAR2013\cite{icdar2013} & \checkmark & \checkmark &\checkmark & 
        PDF & 156 & Heuristics+ML\\ \hline
        ICDAR2019\cite{icdar2019} & \checkmark & \checkmark & $\times$ &
        Images & 3.6k & Heuristics+ML\\ \hline
        UNLV\cite{unlv} & \checkmark & \checkmark & $\times$ & Images & 558 & T-Recs\cite{kieninger1998paper} \\ \hline
        TableBank\cite{tablebank}* & \checkmark&\checkmark & $\times$& Images & 417k (TD) & F-RCNN\cite{frcnn} \\ \hline
        TableBank\cite{tablebank}* & \checkmark&\checkmark & $\times$& Images & 145k (TSR) & WYGIWYS\cite{Deng2016WhatYG}\\ \hline
        SciTSR\cite{scitsr}* & $\times$ & \checkmark & $\checkmark$ & PDF & 15k & GraphTSR\cite{scitsr} \\ \hline
        Table2Latex\cite{table2latex}* & $\times$ & \checkmark & \checkmark & Images & 450k & IM2Tex\cite{im2tex}\\ \hline
        Synthetic data~\cite{graphnn} & $\times$ & \checkmark & \checkmark & Images & Unbounded & DCGNN\cite{graphnn}\\ \hline
        PubTabNet\cite{pubtabnet}* & $\times$ & \checkmark & \checkmark & Images & 568k & EDD\cite{pubtabnet}\\ \hline \hline
         \textsc{TabLeX} (ours)* & $\times$ & \checkmark & \checkmark & Images & 1m\texttt{+} & TRT~\cite{feng2020scene}
        \\ \hline
	\end{tabular}
	}
\end{table}

\noindent \textbf{Scientific tabular datasets:}
Table2Latex~\cite{table2latex} dataset contains 450k scientific table images and its corresponding ground-truth in \LaTeX. It is curated from the \textit{arXiv}\footnote{\label{arxivfootnote}\url{http://arxiv.org/}} repository. To the best of our knowledge, Table2Latex is the only dataset that contains ground truth in \TeX~language but is not publicly available. TableBank~\cite{tablebank} contains 145k table images along with its corresponding ground truth in the HTML representation. TableBank~\cite{tablebank} contains table images from both Word and \LaTeX~documents curated from the internet and \textit{arXiv}\footref{arxivfootnote}, respectively. However, it does not contain content information to perform the TCR task. PubTabNet~\cite{pubtabnet} contains over 568k table images and corresponding ground truth in the HTML language. PubTabNet is curated from the PubMed Central repository\footnote{\label{pubmed}\url{https://pubmed.ncbi.nlm.nih.gov/}}. However, it is only limited to the biomedical and life sciences domain. SciTSR~\cite{scitsr} contains 15k tables in the PDF format with table cell content and the coordinate information in the JSON format. SciTSR has been curated from the \textit{arXiv}\footref{arxivfootnote} repository. However, manual analysis tells that 62 examples out of 1000 randomly sampled examples are incorrect~\cite{TableRobot}.

\noindent \textbf{TIE methodologies:}
The majority of the TIE methods employ encoder-decoder architecture~\cite{table2latex,tablebank,pubtabnet}.  Table2Latex~\cite{table2latex} uses IM2Tex~\cite{im2tex} model where encoder consists of convolutional neural network (CNN) followed by bidirectional LSTM and decoder consists of standard LSTM. TableBank~\cite{tablebank} uses WYGIWS~\cite{Deng2016WhatYG} model, an encoder-decoder architecture, where encoder consists of CNN followed by a recurrent neural network (RNN) and decoder consists of standard RNN. PubTabNet~\cite{pubtabnet} uses a proposed encoder-dual-decoder (EDD)~\cite{pubtabnet} model which consists of CNN encoder and two RNN decoders called structure and cell decoder, respectively. In contrast to the above works, SciTSR~\cite{scitsr} proposed a graph neural network-based extraction methodology. The network takes vertex and edge features as input and computes their representations using graph attention blocks, and performs classification over these edges.

\noindent \textbf{TIE metrics:}
Table2Latex~\cite{table2latex} used BLEU~\cite{bleu} score (a text-based metric) and exact match accuracy (a vision-based metric) for evaluation.
TableBank~\cite{tablebank} also conducted BLEU~\cite{bleu} metric for evaluation.  Tesseract OCR~\cite{smith2007overview} uses the Word Error Rate (WER) metric for evaluation of the output. SciTSR~\cite{scitsr} uses micro- and macro-averaged precision, recall, and F1-score to compare the output against the ground truth, respectively. In contrast to the above standard evaluation metrics in NLP literature,  PubTabNet~\cite{pubtabnet} proposed a new metric called Tree-Edit-Distance-based Similarity (TEDS) for evaluation of the output HTML representation.

\noindent \textbf{The challenges:} Table~\ref{datasets} shows that there are only two datasets for Image-based TCR from scientific tables, that is, Table2Latex~\cite{table2latex} and PubTabNet~\cite{pubtabnet}. We address some of the challenges from previous works with \textsc{TabLeX} which includes (i) large-size for training (ii) Font-agnostic learning (iii) Image-based scientific TCR (iv) domain independent

\section{The \textsc{TabLeX} Dataset}
\label{sec:dataset}
This section presents the detailed curation pipeline to create the \textsc{TabLeX} dataset. Next, we discuss the data acquisition strategy.  


\subsection{Data Acquisition}
\begin{figure}[!t]
\centering
\resizebox{\hsize}{!}{
\includegraphics[]{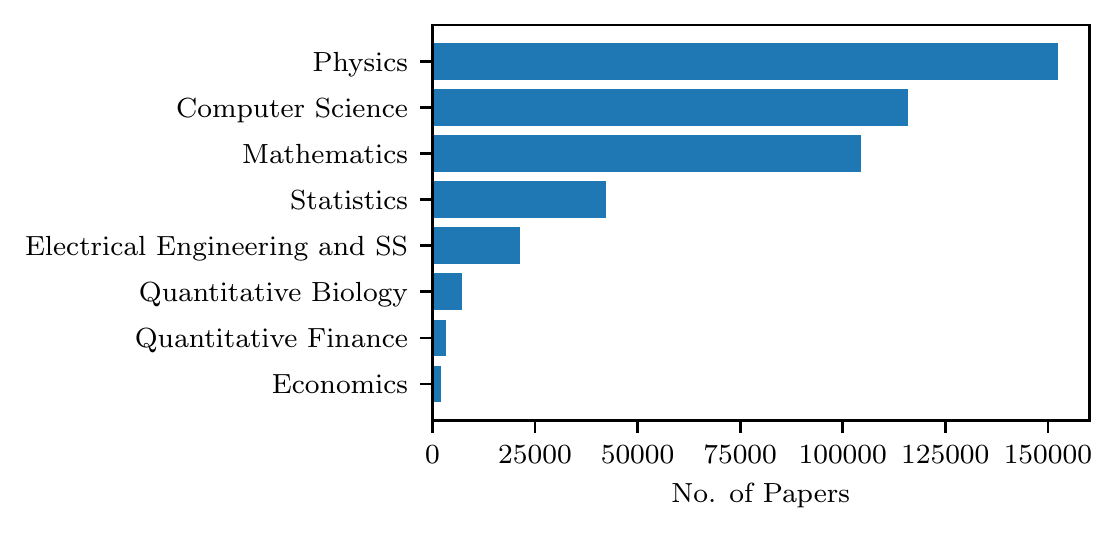}
}
\caption{Total number of papers in arXiv's subject categories. Here SS denotes Systems Science.}
\label{fig:arxiv_freq}
\end{figure}

We curate data from popular preprint repository \textit{arXiv}\footref{arxivfootnote}.
We downloaded paper (uploaded between Jan 2019--Sept 2020) source code and corresponding compiled PDF. These articles belong to eight subject categories. Figure~\ref{fig:arxiv_freq} illustrates category-wise paper distribution. As illustrated, the majority of the papers belong to three subject categories, physics (33.93\%), computer science (25.79\%), and mathematics (23.27\%). Overall, we downloaded 347,502 papers and processed them using the proposed data processing pipeline (described in the next section).  

\subsection{Data Processing Pipeline}
The following steps present a detailed description of the data processing pipeline.
\begin{figure}[!tbh]
\setlength{\tabcolsep}{1pt}
\begin{tabular}{cc}
    \includegraphics[scale=0.3]{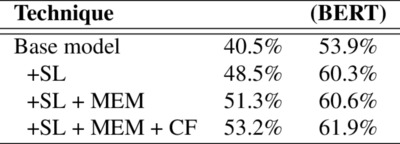} & \includegraphics[scale=0.52]{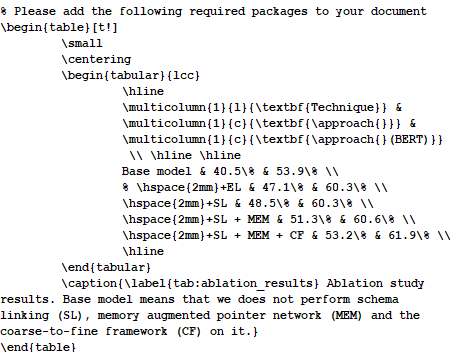} \\
    (a) A real table image & (b) Corresponding \LaTeX~source code\\
    \includegraphics[scale=0.55]{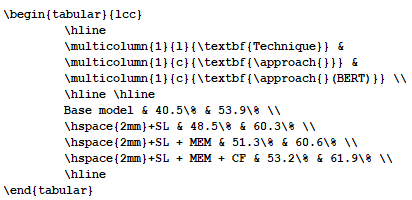} & \includegraphics[scale=0.65]{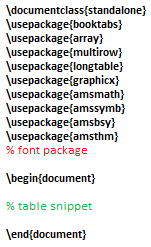}\\
     (c) Extracted table snippet & (d) \LaTeX~code template\\
     \includegraphics[scale=0.55]{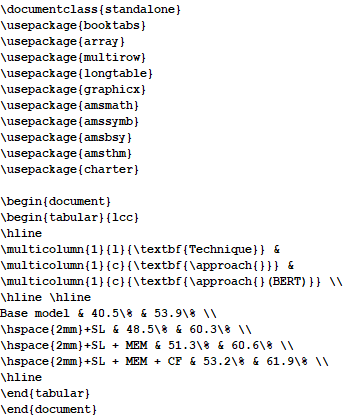} & \includegraphics[scale=0.3]{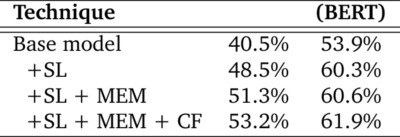}\\
      (e) Code with \textsc{Charter} font package& (f) The final generated table image \\
\end{tabular}
    \caption{Data processing pipeline.}
    \label{fig:pipeline}
\end{figure}

\subsubsection{\LaTeX~Code Pre-Processing Steps}
\begin{enumerate}
    \item \textbf{Table Snippets Extraction:} A table snippet is a part of \LaTeX~code that begins with `\verb+\begin{tabular}+' and ends with `\verb+\end{tabular}+' command. During its extraction, we removed citation command `\verb+\cite{}+', reference command `\verb+\ref{}+', label command `\verb+\label{}+', and graphics command `\verb+\includegraphics[]{}+', along with $\sim$ symbol (preceding these commands). Also, we remove  command pairs (along with the content between them) like  `\verb+\begin{figure}+' and `\verb+\end{figure}+', and `\verb+\begin{subfigure}+' and `\verb+\end{subfigure}+', as they cannot be predicted from the tabular images. Furthermore, we also remove the nested table environments. Figure~\ref{fig:pipeline}a and \ref{fig:pipeline}b show an example table and its corresponding \LaTeX~source code, respectively. 
    \item \textbf{Comments Removal:} Comments are removed by removing the characters between `\verb+%+' token and newline token `\verb+\n+'. This step was performed because comments do not contribute to visual information. 
    \item \textbf{Column Alignment and Rows Identification:} We keep all possible alignment tokens (`\verb+l+', `\verb+r+', and `\verb+c+') specified for column styling. The token `\verb+|+' is also kept to identify vertical lines between the columns, if present. The rows are identified by keeping the `\verb+\\+' and `\verb+\tabularnewline+' tokens. These tokens signify the end of each row in the table. For example, Figure~\ref{fig:pipeline}c shows extracted table snippet from \LaTeX~source code (see Figure~\ref{fig:pipeline}b) containing the column and rows tokens with comment statements removed.
    \item \textbf{Font Variation:} In this step, the extracted \LaTeX~code is augmented with different font styles. We experiment with a total 12 different \LaTeX~font packages\footnote{\url{https://www.overleaf.com/learn/latex/font_typefaces}}. We use popular font packages from PostScript family which includes `\verb+courier+', `\verb+helvet+', `\verb+palatino+', `\verb+bookman+', `\verb+mathptmx+', `\verb+utopia+' and also other font packages such as `\verb+tgbonum+', `\verb+tgtermes+', `\verb+tgpagella+', `\verb+tgschola+', `\verb+charter+ and `\verb+tgcursor+'. For each curated image, we create 12 variations, one in each of the font style. 
    \item \textbf{Image Rendering:} Each table variant is compiled into a PDF using a \LaTeX~code template (described in Figure~\ref{fig:pipeline}d). Here, \textit{`table snippet'} represents the extracted tabular \LaTeX~code and \textit{`font package'} represents the \LaTeX~font package used to generated the PDF. The corresponding table PDF files are then converted into JPG images using the Wand library\footnote{\url{https://github.com/emcconville/wand}} which uses ImageMagick~\cite{imagemagick} API to convert PDF into images. Figure~\ref{fig:pipeline}e shows the embedded code and font information within the template. Figure~\ref{fig:pipeline}f shows the final table image with \textsc{Charter} font. During conversion, we kept the image's resolution as 300 dpi, set the background color of the image to white, and removed the transparency of the alpha channel by replacing it with the background color of the image. We use two types of aspect ratio variations during the conversion, described as follows:
    \begin{enumerate}
    \item \textit{Conserved Aspect Ratio:}
    In this case, the bigger dimension (height or width) of the image is resized to 400 pixels\footnote{We use `400' pixels as an experimental number.}. We then resize the smaller dimension (width or height) by keeping the original aspect ratio conserved. During resizing, we use a blur factor of 0.8 to keep images sharp. 
    \item \textit{Fixed Aspect Ratio:}
    The images are resized to a fixed size of 400$\times$400 pixels using a blur factor of 0.8. Note that this resizing scheme can lead to extreme levels of image distortions. 
\end{enumerate}
    
\end{enumerate}

\subsubsection{\LaTeX~Code Post-Processing Steps for Ground-Truth Preparation}
\begin{enumerate}
    \item \textbf{Filtering Noisy \LaTeX~tokens:} We filter out \LaTeX~environment tokens (tokens starting with `\verb+\+') having very less frequency (less than 5000) in the overall corpus compared to other \LaTeX~environment tokens and replaced it with `\verb+\LATEX_TOKEN+' as these tokens will have a little contribution. This frequency-based filtering reduced the corpus's overall vocabulary size, which helps in training models on computation or memory-constrained environments. 
    \item \textbf{Structure Identification:} In this post-processing step, we create Table Structure Dataset (TSD). It consists of structural information corresponding to the table images. We keep tabular environment parameters as part of structure information and replace tokens representing table cells' content with a placeholder token `CELL'. This post-processing step creates ground truth for table images in TSD. Specifically, the vocabulary comprises digits (0-9), `\&',  `CELL', `\verb+\\+', `\verb+\hline+', `\verb+\multicolumn+',  `\verb+\multirow+', `\verb+\bottomrule+', `\verb+\midrule+', `\verb+\toprule+', `\verb+|+', `l', `r', `c', `\{', and `\}'. A sample of structure information is shown in Figure~\ref{fig:table_example}b, where the `CELL' token represents a cell structure in the table. Structural information can be used to identify the number of cells, rows, and columns in the table using `CELL', \verb+`\\'+ and alignment tokens (`c', `l', and `r'), respectively. Based on output sequence length, we divided this dataset further into two variants TSD-250 and TSD-500, where the maximum length of the output sequence is 250 and 500 tokens, respectively. 
    \item \textbf{Content Identification:} Similar to TSD, we also create a Table Content Dataset (TCD). This dataset consists of content information including alphanumeric characters, mathematical symbols, and other \LaTeX~environment tokens. Content tokens are identified by removing tabular environment parameters and keeping only the tokens that identify table content. Specifically, the vocabulary includes all the alphabets (a-z and A-Z), digits (0-9), \LaTeX~environment tokens (`\verb+\textbf+', \verb+\hspace+', etc.), brackets (`\verb+(+', `\verb+)+', `\verb+{+', `\verb+}+', etc.) and all other possible symbols (`\verb+$+', `\verb+&+', etc.). In this dataset, based on output sequence length, we divided it further into two variants TCD-250 and TCD-500, where the maximum length of the output sequence is 250 and 500 tokens, respectively. A sample of content information is shown in Figure~\ref{fig:table_example}c.
\end{enumerate}

\subsection{Dataset Statistics}
\begin{figure}[h]
\begin{subfigure}[h]{.5\textwidth}
\centering
\resizebox{\textwidth}{!}{
\includegraphics{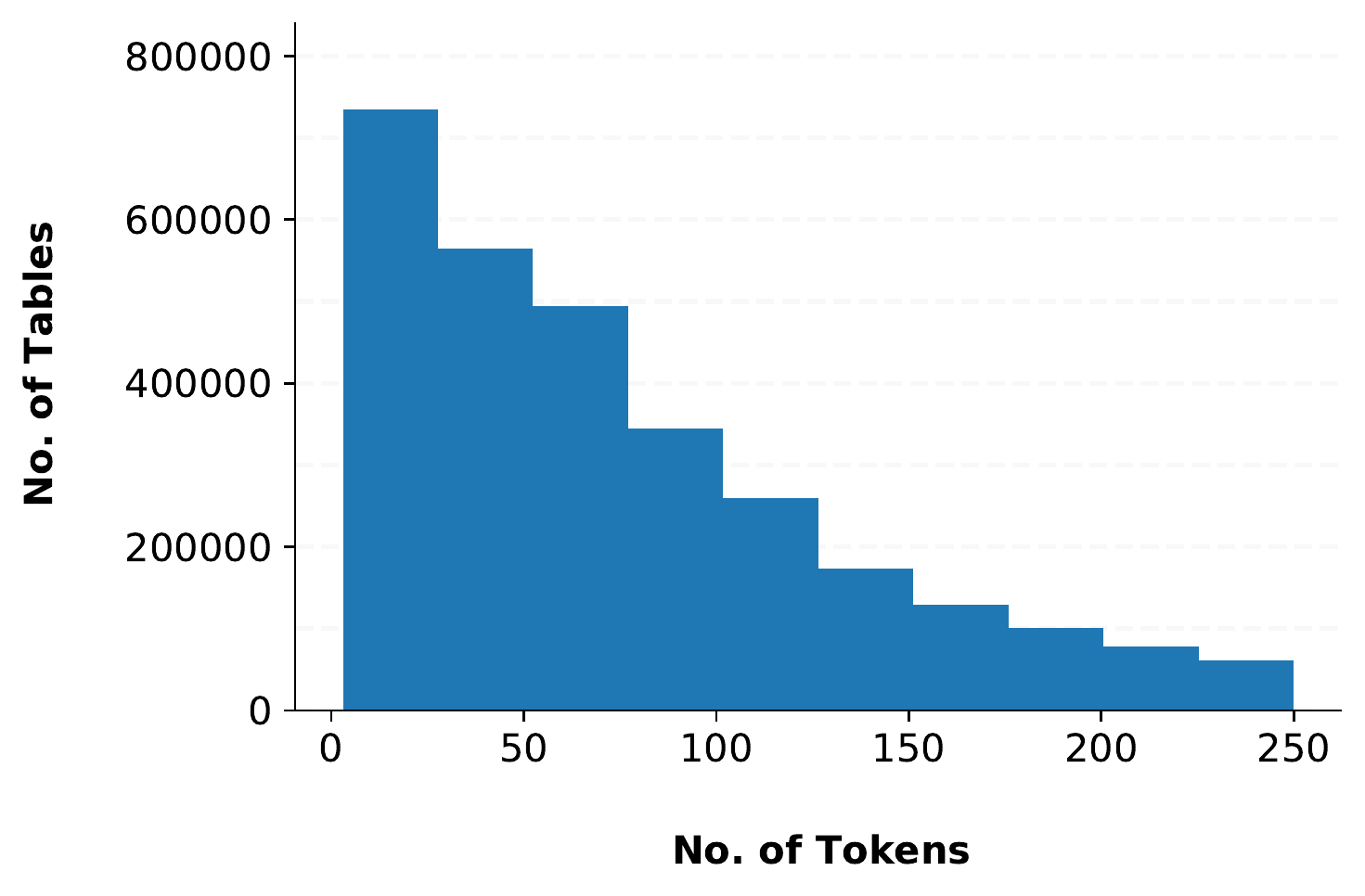}
}
\caption{TSD-250}
\end{subfigure}
\hfill
\begin{subfigure}[h]{.5\textwidth}
\centering
\resizebox{\textwidth}{!}{
\includegraphics{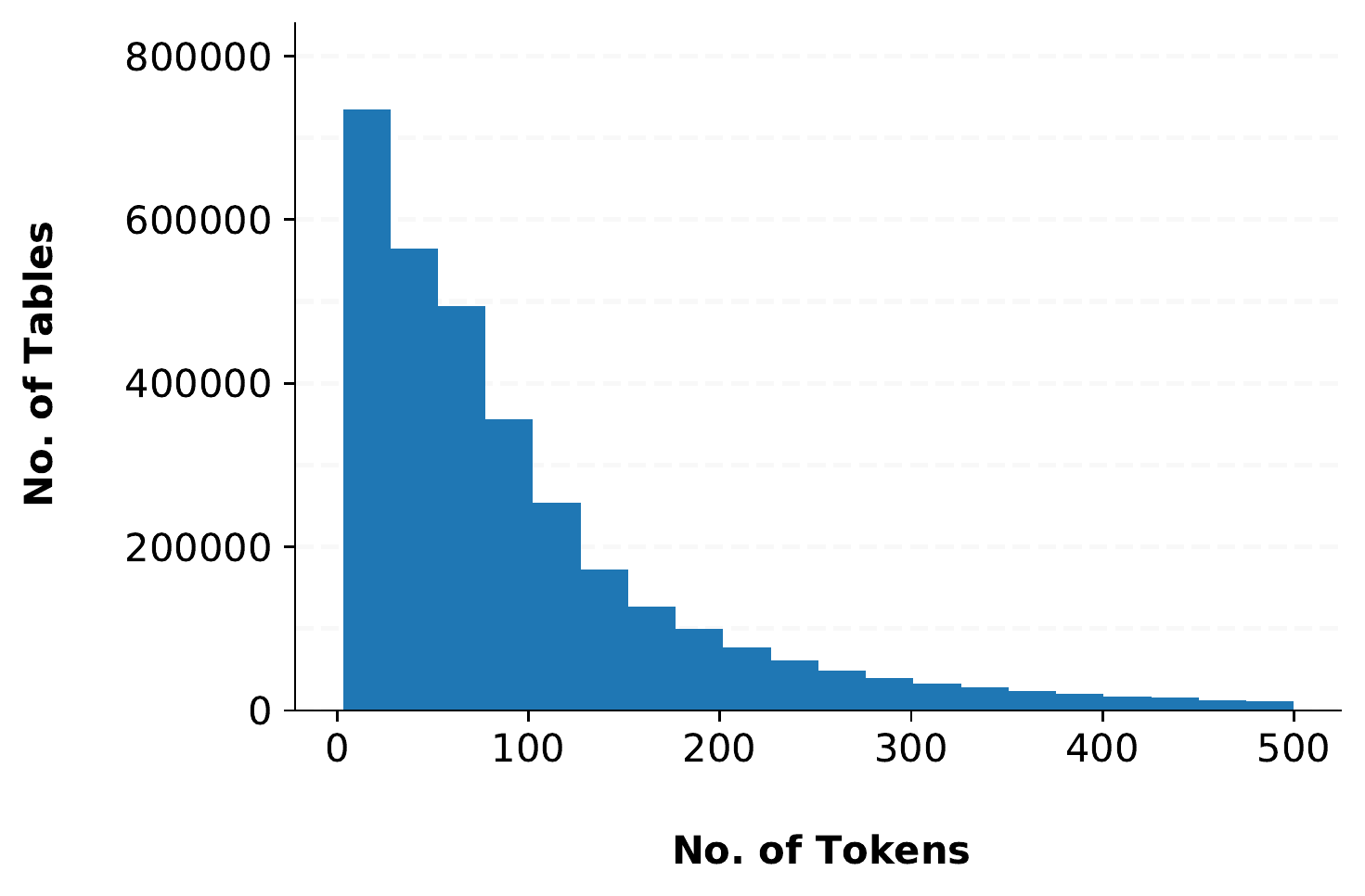}
}
\caption{TSD-500}
\end{subfigure}
\hfill
\begin{subfigure}[h]{.5\textwidth}
\centering
\resizebox{\textwidth}{!}{
\includegraphics{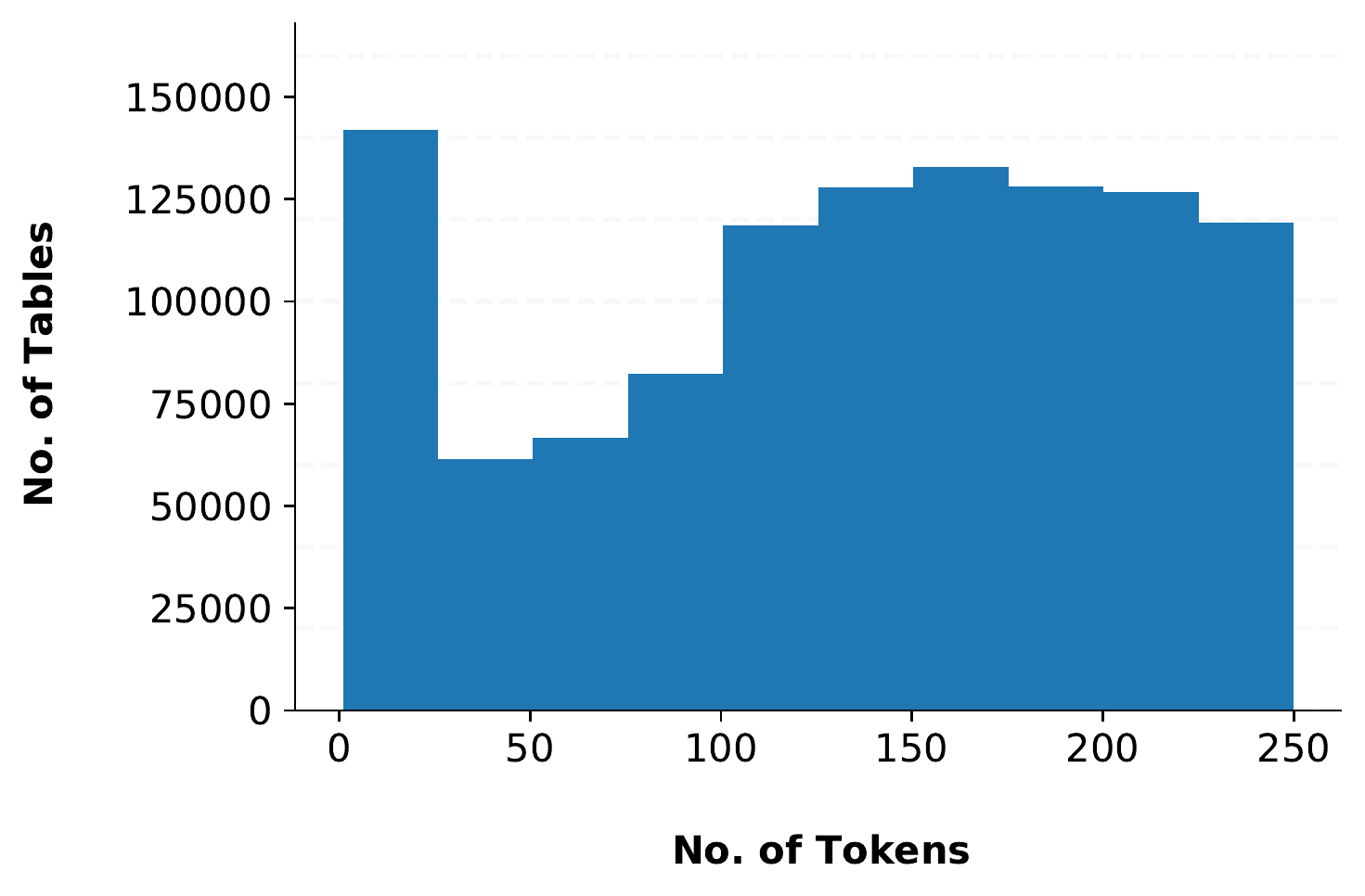}
}
\caption{TCD-250}
\end{subfigure}
\begin{subfigure}[h]{.5\textwidth}
\centering
\resizebox{\textwidth}{!}{
\includegraphics[scale=0.65]{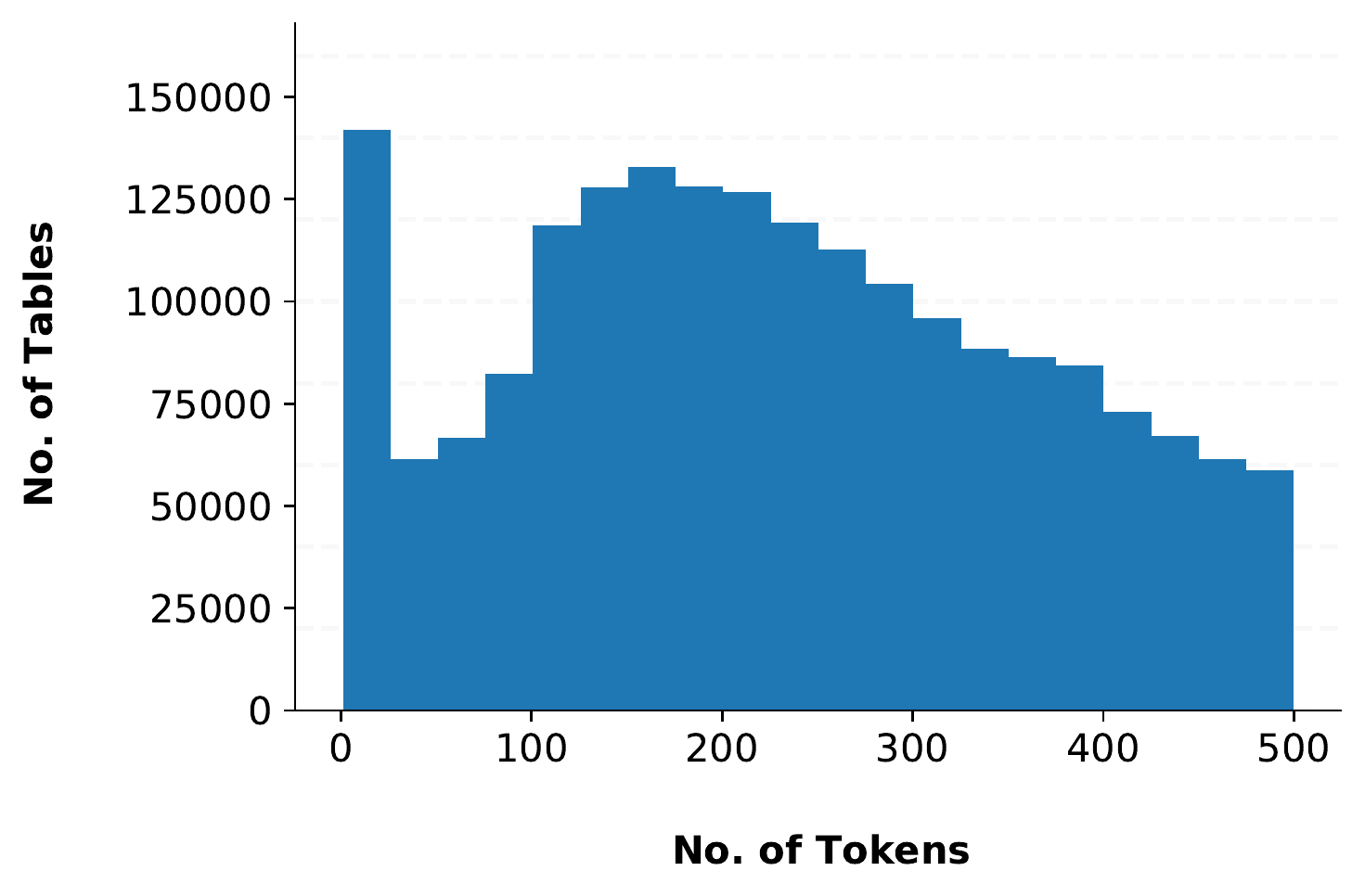}
}
\caption{TCD-500}
\end{subfigure}
\caption{Histograms representing the number of tokens distribution for the tables present in the dataset.}
\label{histograms}
\end{figure}

We further partition each of the four dataset variants TSD-250, TSD-500, TCD-25, and TCD-500, into training, validation, and test sets in a ratio of 80:10:10. Table~\ref{tab_datasets} shows the summary of the dataset. Note that the number of samples present in the TSD and the corresponding TCD can differ due to variation in the output sequence's length.  Also, the average number of tokens per sample in TCD is significantly higher than the TSD due to more information in the tables' content than the corresponding structure. Figure~\ref{histograms} demonstrates histograms representing the token distribution for the TSD and TCD. In the case of TSD, the majority of tables contain less than 25 tokens. Overall the token distribution shows a long-tail behavior. In the case of TCD, we witness a significant proportion between 100--250 tokens. The dataset is licensed under \href{https://creativecommons.org/licenses/by-nc-sa/4.0/}{Creative Commons Attribution 4.0 International License} and available for download at \url{https://www.tinyurl.com/tablatex}.

\begin{table}[!t]
\caption{\textsc{TabLeX} Statistics. ML denotes the maximum length of output sequence. Samples denotes the total number of table images. T/S denotes the average number of tokens per sample. VS denotes the vocabulary size. AR and AC denote the average number of rows and columns among all the samples in the four dataset variants, respectively.}
\label{tab_datasets}
\centering
\begin{tabular}{|l|c|c|c|c|c|c|c|}
		\hline
        \multicolumn{1}{|l|}{\textbf{Dataset}} &
		\multicolumn{1}{|c}{\textbf{ML}} &
		\multicolumn{1}{|c}{\textbf{Samples}} &
        \multicolumn{1}{|c}{\textbf{Train/Val/Test}} &
        \multicolumn{1}{|c|}{\textbf{T/S}} &
        \multicolumn{1}{|c|}{\textbf{VS}} &
        \multicolumn{1}{|c|}{\textbf{AR}} & 
        \multicolumn{1}{|c|}{\textbf{AC}} \\ \hline
        TSD-250 & 250 & \num{2938392}  & \num{2350713} / \num{293839} / \num{293840} & \num{74.09} & 25 & 6 & 4\\ \hline
        TSD-500 & 500 & \num{3191891}  & \num{2553512} / \num{319189} / \num{319190} & \num{95.39} & 25 & 8 & 5\\ \hline
  		TCD-250 & 250 & \num{1105636} & \num{884508} / \num{110564} / \num{110564} & 131.96 & 718 & 4 & 4\\ \hline
  		TCD-500 & 500 & \num{1937686} & \num{1550148} / \num{193769} / \num{193769} & \num{229.06} & 737 & 5 & 4\\ \hline
	\end{tabular}
\end{table}

\section{Evaluation Metrics}
\label{sec:metrics}
\begin{figure}
    \centering
    \subfloat[\centering A Toy Table ]{{\includegraphics[scale=0.95]{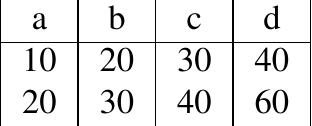} }}%
    \label{a}
    \qquad
    \subfloat[\centering Ground Truth ]{{\includegraphics[scale=0.7]{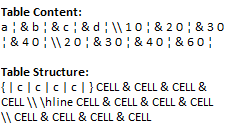} }}%
    \qquad
    \subfloat[\centering Table Content Output ]{{\includegraphics[scale=0.7]{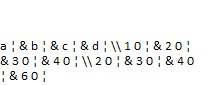} }}%
    \qquad
    \subfloat[\centering Table Structure Output ]{{\includegraphics[scale=0.7]{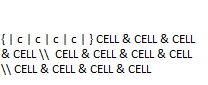} }}%
    \caption{(a) A toy table, (b) corresponding ground truth token sequence for TSR and TCR tasks, (c) model output sequence for TCR task, and (d) model output sequence for TSR task.}
    \label{fig:dummy_table}%
\end{figure}

We experiment with three evaluation metrics to compare the predicted sequence of tokens ($T_{PT}$) against the sequence of tokens in the ground truth ($T_{GT}$). Even though the proposed metrics are heavily used in NLP research, few TIE works have leveraged them in the past. Next, we discuss these metrics and illustrate the implementation details using a toy example described in Figure~\ref{fig:dummy_table}.

\begin{enumerate}
    \item \noindent\textbf{Exact Match Accuracy (EMA):} EMA outputs the fraction of predictions with exact string match of $T_{PT}$ against $T_{GT}$. A model having a higher value of EMA is considered a good model. In Figure~\ref{fig:dummy_table}, for the TSR task, $T_{PT}$ misses `\verb+\hline+' token compared to the corresponding $T_{GT}$, resulting in EMA = 0. Similarly, the $T_{PT}$ exactly matches the $T_{GT}$ for the TCR task, resulting in EMA = 1.   
    
    \item \noindent\textbf{Bilingual Evaluation Understudy Score (BLEU):}   
  The BLEU~\cite{bleu} score is primarily used in Machine Translation literature to compare the quality of the translated sentence against the expected translation.  Recently, several TIE works~\cite{Deng2016WhatYG,im2tex} have adapted BLEU for evaluation. It counts the matching n-grams in $T_{PT}$  against the n-grams in $T_{GT}$.  The comparison is made regardless of word order. The higher the BLEU score, the better is the model. We use SacreBLEU~\cite{sacrebleu} implementation for calculating the BLEU score. We report scores for the most popular variant, BLEU-4. BLEU-4 refers to the product of brevity penalty (BP) and a harmonic mean of precisions for unigrams, bigrams, 3-grams, and 4-grams (for more details see~\cite{bleu}). Figure~\ref{fig:dummy_table}, there is an exact match between $T_{PT}$ and $T_{GT}$ for TCR task yielding BLEU = 100.00. In the case of TSR, the missing `\verb+\hline+' token in $T_{PT}$ yields a lower BLEU = 89.66.  
    
    \item \noindent\textbf{Word Error Rate (WER):} WER is defined as a ratio of the Levenshtein distance between the $T_{PT}$ and $T_{GT}$ to the length of $T_{GT}$.  It is a standard evaluation metric for several OCR tasks. Since it measures the rate of error, models with lower WER are better. We use jiwer\footnote{\url{https://github.com/jitsi/jiwer}} Python library for WER computation. In Figure~\ref{fig:dummy_table}, for the TCR task, the WER between $T_{PT}$ and $T_{GT}$ is 0. Whereas in the case of TSR, the Levenshtein distance between $T_{PT}$ and $T_{GT}$ is one due to the missing `\verb+\hline+' token. Since the length of $T_{GT}$ is 35, WER comes out to be 0.02. 

\end{enumerate}

\section{Experiments}
\label{sec:exp}
\begin{figure}[h!]
\centering
\includegraphics[scale=0.6]{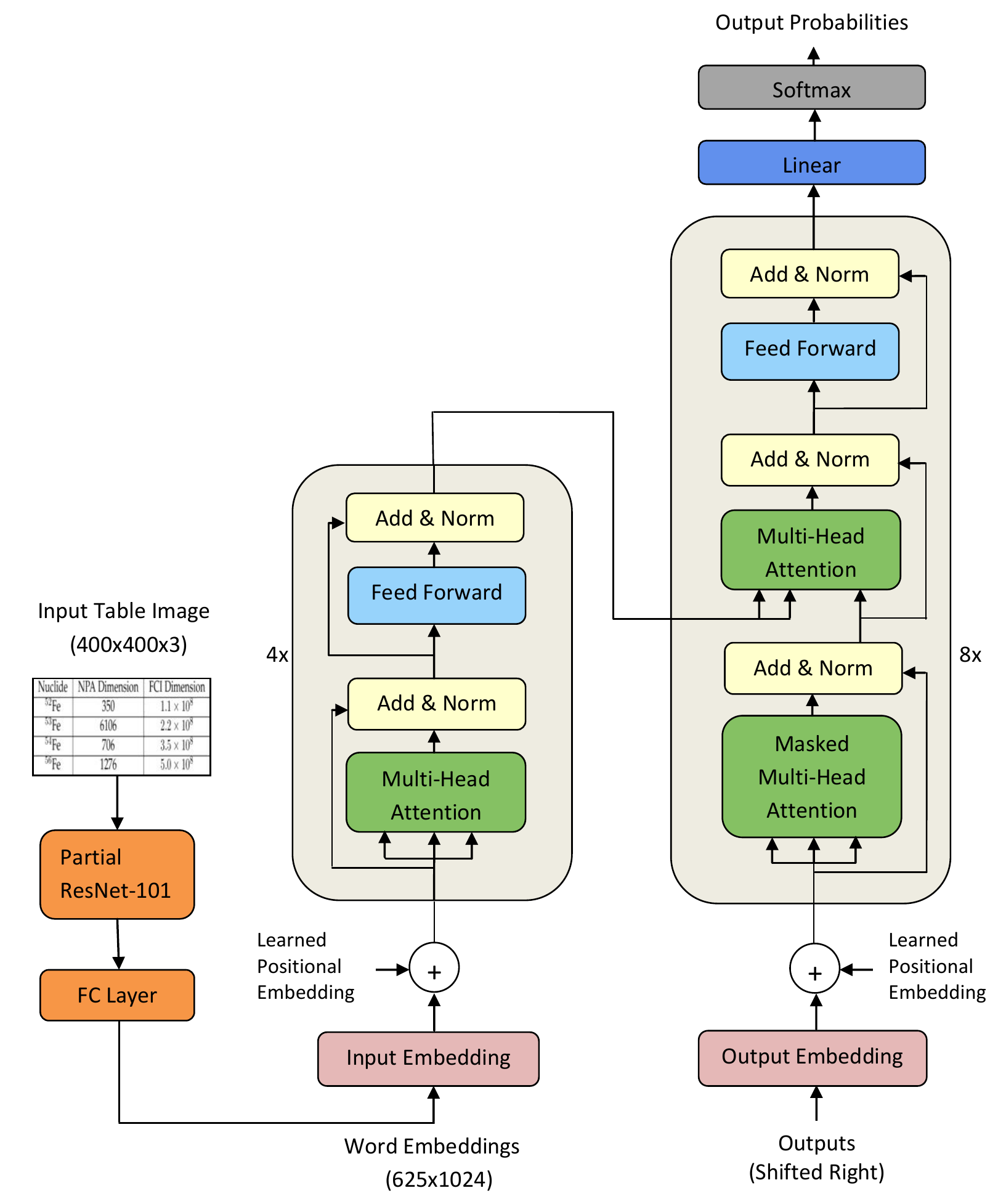}
\caption{TIE-ResNet-Transformer model architecture.}
\label{fig:baseline}
\end{figure}

In this section, we experiment with a deep learning-based model for TSR and TCR tasks. We adapt an existing model~\cite{feng2020scene} architecture proposed for the scene text recognition task and train it from scratch on the \textsc{TabLeX} dataset. It uses partial ResNet-101 along with a fully connected layer as a feature extractor module for generating feature embeddings with a cascaded Transformer~\cite{vaswani2017attention} module for encoding the features and generating the output. Figure~\ref{fig:baseline} describes the detailed architecture of the model. Note that in contrast to the scene image as an input in ~\cite{feng2020scene}, we input a tabular image and predict the \LaTeX~token sequence. We term it as the TIE-ResNet-Transformer model (\textit{TRT}).

\subsection{Implementation Details} 
For both TSR and TCR tasks, we train a similar model. We use the third intermediate hidden layer of partial ResNet-101 and an FC layer as a feature extractor module to generate a feature map. The generated feature map (size = $625\times1024$) is further embedded into 256 dimensions, resulting in reduced size of $625\times256$. We experiment with four encoders and eight decoders with learnable positional embeddings. The models are trained for ten epochs with a batch size of 32 using cross-entropy loss function and Adam Optimizer with an initial learning rate of $0.1$ and 2000 warmup training steps, decreasing with Noam's learning rate decay scheme. A dropout rate of 0.1 and $\epsilon_{ls}=0.1$ label smoothing parameter is used for regularization. For decoding, we use the greedy decoding technique in which the model generates an output sequence of tokens in an auto-regressive manner, consuming the previously generated tokens to generate the next token. All the experiments were performed on 4$\times$NVIDIA V100 graphics card. 
\subsection{Results}

\begin{table}[h]
    \caption{TRT results on the \textsc{TabLeX} dataset.}
	\label{results}
	\centering
	\resizebox{.7\hsize}{!}{
    \begin{tabular}{| c | c || c | c | c | }
        \hline
        Dataset & Aspect Ratio & EMA(\%) & BLEU & WER(\%) \\ \hline

    \multirow{2}{*}{\makecell{TCD \\ 250}} & Conserved & 21.19 & 95.18 & 15.56  \\
         \cline{2-5}
        
     & Fixed & 20.46 & 96.75 & 14.05  
     \\ \hline
        
    \multirow{2}{*}{\makecell{TCD\\ 500}} & Conserved  & 11.01 & 91.13 & 13.78 \\
            \cline{2-5}
            
    & Fixed & 11.23 & 94.34 & 11.12  \\
        \hline\hline
  
    \multirow{2}{*}{\makecell{TSD \\ 250}}  & Conserved  & 70.54 & 74.75 & 3.81 \\
         \cline{2-5}
        
    & Fixed & 74.02 & 70.59 & 4.98  \\ 
    \hline
        
    \multirow{2}{*}{\makecell{TSD\\ 500}}  & Conserved  & 70.91 & 82.72 & 2.78 \\ 
            \cline{2-5}
            
    & Fixed & 71.16 & 61.84 & 9.34  \\ 
         \hline
    \end{tabular}
    }
\end{table}

Table~\ref{results} illustrates that higher sequence length significantly degrades the EMA score for both the TSR and TCR tasks. For TCD-250 and TCD-500, high BLEU scores ($>$90) suggest that the model can predict a large chunk of \LaTeX~content information correctly. However, the errors are higher for images with a conserved aspect ratio than with a fixed aspect ratio, which is confirmed by comparing their BLEU score and WER metrics. Similarly, for TSD-250 and TSD-500, high EMA and lower WER suggest that the model can correctly identify structure information for most of the tables. 
TCD yields a lower EMA score than TSD. We attribute this to several reasons. One of the reasons is that the TCR model fails to predict some of the curly braces (`\{' and `\}') and dollar (`\$') tokens in the predictions. After removing curly braces and dollar tokens from the ground truth and predictions, EMA scores for conserved and fixed aspect ratio images in TCD-250 increased to 68.78\% and 75.33\%, respectively. Similarly, for TCD-500, EMA for conserved and fixed aspect ratio images increases to 49.23\% and 59.94\%, respectively. In contrast, TSD do not contain dollar tokens, and curly braces are only present at the beginning of the column labels, leading to higher EMA scores. In the future, we believe the results can be improved significantly by proposing better vision-based DL architectures for TSR and TCR tasks.

\section{Conclusion}
\label{section6}
This paper presents a benchmark dataset \textsc{TabLeX}  for structure and content information extraction from scientific tables. It contains tabular images in 12 different fonts with varied visual complexities. We also proposed a novel preprocessing pipeline for dataset generation. We evaluate the existing state-of-the-art transformer-based model and show excellent future opportunities in developing algorithms for IE from scientific tables. In the future, we plan to continuously augment the dataset size and add more complex tabular structures for training and prediction. 

\section*{Acknowledgment}
This work was supported by The Science and
Engineering Research Board (SERB), under sanction number ECR/2018/000087.

\bibliographystyle{splncs04}
\bibliography{sample}
\end{document}